\documentclass[fleqn,usenatbib]{mnras}

\usepackage{newtxtext,newtxmath}

\usepackage[T1]{fontenc}

\DeclareRobustCommand{\VAN}[3]{#2}
\let\VANthebibliography\thebibliography
\def\thebibliography{\DeclareRobustCommand{\VAN}[3]{##3}\VANthebibliography}

\usepackage{graphicx}
\usepackage{float} 
\usepackage{amsmath}

\usepackage{amssymb}	
\usepackage{color}

\usepackage{ulem}

\title[Mapping the Milky Way's stellar halo with 2D data]{Mapping the Milky Way's stellar halo with 2D data}

\author[A. Chen et al.]{
Anda Chen$^{1,2}$
Zhigang Li$^{3}$
Yougang Wang$^{4,5}$\thanks{E-mail: wangyg@bao.ac.cn}
Yan Gong$^{1,6}$
Xuelei Chen$^{4,2,7,5}$
and Richard J. Long$^{8,9}$
\\
$^{1}$Key Laboratory of Space Astronomy and Technology, National Astronomical Observatories, Chinese Academy of Sciences, Beijing 100101, China\\
$^{2}$School of Astronomy and Space Sciences, University of Chinese Academy of Sciences, Beijing 100049, China\\
$^{3}$ College of Physics and Electronic Engineering, Nanyang Normal University, Nanyang, Henan, 473061, China\\
$^{4}$National Astronomical Observatories, Chinese Academy of Sciences, Beijing 100101, China\\
$^{5}$Key Laboratory of Radio Astronomy and Technology, Chinese Academy of Sciences, A20 Datun Road, Chaoyang District, Beijing, 100101, P. R. China\\
$^{6}$Science Center for China Space Station Telescope, National Astronomical Observatories, Chinese Academy of Sciences, Beijing 100101, China\\
$^{7}$Key Laboratory of Cosmology and Astrophysics (Liaoning) \& Department of Physics, College of Sciences, Northeastern University, Shenyang 110819, China\\
$^{8}$Department of Astronomy, Tsinghua University, Beijing 100084, China\\
$^{9}$Jodrell Bank Centre for Astrophysics, Department of Physics and Astronomy, The University of Manchester, Oxford Road, Manchester M13 9PL, UK
}

\date{Accepted XXX. Received YYY; in original form ZZZ}

\pubyear{2023}

\begin{document}
\label{firstpage}
\pagerange{\pageref{firstpage}--\pageref{lastpage}}
\maketitle

\begin{abstract}
We propose a new method for measuring the spatial density distribution of the stellar halo of the Milky Way. Our method is based on a pairwise statistic of the distribution of stars on the sky, the angular two-point correlation function (ATPCF). The ATPCF utilizes two dimensional data of stars only and is therefore immune to the large uncertainties in the determination of distances to stars. We test our method using mock stellar data coming from various models including the single power-law (SPL) and the broken power-law (BPL) density profiles. We also test the influence of axisymmetric flattening factors using both constant and varying values. We find that the ATPCF is a powerful tool for recovering the spatial distributions of the stellar halos in our models. We apply our method to observational data from the type ab RR Lyrae catalog in the Catalina Survey Data Release 1. 
In the 3-parameter BPL model, we find that $s_{1}=2.46_{-0.20}^{+0.18}, s_{2}=3.99_{-1.33}^{+0.75}$ and $r_{0}=31.11_{-5.88}^{+7.61}$, which are in good agreement with previous results. We also find that introducing an extra parameter, the radially varying flattening factor, greatly improves our ability to model accurately the observed data distribution. This implies perhaps that the stellar halo of the Milky Way should be regarded as oblate. 
\end{abstract}




\section{Introduction}

The stellar halo is an important component of the Milky Way (MW). Although it contains only a small fraction ($\sim 1\%$) of the stellar mass of the Galaxy (e.g., \citealt{Xue_2015,Hernitschek2018}), its spatial distribution is crucial for understanding the formation and evolution of our galaxy. According to the theory of hierarchical structure formation, during the process of galaxy evolution, the growth of a galaxy is achieved by mergers and the accretion of smaller systems. Since stellar halos formed from disrupted satellites, the relics of these events may still remain, especially in the outermost regions, as the long dynamical timescales of galaxies allow accretion related substructures to linger for gigayears. Studies of the Galactic stellar halo will help us to understand the formation history of our Galaxy. 

The stellar density distribution, as well as the metallicities, kinematics, and the ages of halo stars, enables us to obtain information about these important merger events. \cite{Deason2014ApJ} used A-type stars when comparing with the numerical simulations from \cite{Bullock2005ApJ}, and showed that stellar halos with shallower slopes at large distances tend to have more recent accretion activity. Also, as demonstrated by \cite{Deason2018ApJ}, the `break' that might exist in the Galactic halo's radial density profile, can be caused by tidal debris from a merging satellite when it is at the apocentre. \cite{Monachesi2019MNRAS} examined the stellar halos of the Auriga simulations (\citealt{Grand2017MNRAS}). The halos in this simulation have different masses and density profiles, mean metallicity and metallicity gradients, ages, and shapes, reflects the stochasticity inherent in their accretion and merger histories. Their results showed that galaxies with few significant progenitors have more massive halos have steeper density profiles, and possess large negative halo metallicity gradients. Using a cumulative “close pair distribution” as a statistic in  phase space, \cite{Xue2011ApJ} quantified the presence of position–velocity substructure at high statistical significance among the Milky Way halo's BHB stars from SDSS DR8. The comparison between their results with 11 halo formation simulations confirmed the hierarchical build-up of the stellar halo through a signature in phase space. Using data from APOGEE and Gaia, \cite{Horta2023MNRAS} examined the chemical properties of different halo substructures, and classified them through their different possible origins in the assembly history of the Galaxy. \cite{Dodd2023A&A} utilised a single linkage-based clustering algorithm to identify overdensities in the integrals of motion space which may be caused by merger debris. Combined with metallicity information and chemical abundances, they found that the local stellar halo contains an important amount of substructure originating both in situ and from accretion. 

A simple and much explored approach is to measure the number density profile of the stellar halo by star counts.  Over the past few decades, a number of works on stellar halo have been carried out, but a consensus on its shape and radial profile is yet to be reached. Early studies usually described the Galactic halo with an oblate single power law (SPL) model, with a power index of
$2.5-3.5$ and a constant flattening $q$ in the range of $0.5-1$ \citep{Harris1976AJ,Hawkins1984MNRAS,Sommer-Larsen1987MNRAS,Preston1991ApJ,Soubiran1993A&A,Wetterer1996AJ,Gould1998ApJ,Reid1998ASPC,Vivas2006}. However, there is also evidence that a single power law may not be adequate. An early example is \cite{Saha1985ApJ} where they used RR lyrae stars and found that the stellar halo follows a broken power law (BPL) distribution with a power index of $3$ when the radius is less than $25$ kpc, while for greater radii the index is $5$. Some other researchers imply that the distribution may be even more complicated. For example, \cite{Preston1991ApJ} found that the stellar halo becomes more spherical as the Galactic radius increases. 

Large sky surveys, such as the Sloan Digital Sky Survey (PDSS I-III; \citealt{York2000AJ,Eisenstein2011AJ}), the Two-Micron All Sky Survey (2MASS; \citealt{Skrutskie2006AJ}), the Radial Velocity
Experiment (RAVE; \citealt{Steinmetz2020AJ}), the Large Sky Area Multi-object Fiber Spectroscopic Telescope (LAMOST; \citealt{Cui2012RAA,Zhao2012RAA,Deng2012RAA}), Catalina (\citealt{Drake2014ApJS}), Gaia (\citealt{Gaia2016A&A}), the Pan-STARRS1 (PS1;\citealt{Kaiser2010SPIE}), have triggered major growth in this field of research in the past two decades, and provided more samples with greater depth and larger sky coverage. While the SPL model with a constant flattening factor gives a simple description of the radial distribution of the MW's stellar halo \citep{Robin2000,Siegel2002ApJ,Bell2008,Juric2008ApJ}, the BPL model with a constant flattening can fit the data better sometimes \citep{Deason2011,Sesar2011,Pila-Diez2015}. Variable flattening has also been introduced. \cite{Xue_2015} demonstrated that an SPL model with a radius-dependent flattening factor can fit the observational stellar halo well, as do BPL or Einasto profiles \citep{Einasto1965}. Using a non-parametric method, \cite{Xu2018} found that the number density profile is well described by an SPL model with power-law index $5.03_{-0.64}^{+0.64}$, and a radial variable flattening factor $q$, with values about 0.64, 0.8 and 0.96 at $r = 15, 20$ and $30$ kpc. \cite{Hernitschek2018} applied a forward modelling technique, and studied several density profile models, including the SPL, BPL and Einasto models. For each model, they also considered constant and variable flattening. In previous studies, different tracers have been adopted, such as globular clusters \citep{Harris1976AJ}, RR Lyrae variables \citep{Hawkins1984MNRAS,Saha1985ApJ,Wetterer1996AJ,Vivas2006,Hernitschek2018,Iorio2018MNRAS,Li2022MNRAS}, K giant stars \citep{Xue_2015,Xu2018}, blue horizontal branch (BHB) stars \citep{Sommer-Larsen1987MNRAS,Deason2011} and K dwarfs \citep{Gould1998ApJ}. Some studies also used more than one tracer. For example, a combined dataset of RR Lyrae and BHB stars is used by \citep{Preston1991ApJ}. 
 
The two point correlation function (TPCF) is one of the most commonly used statistics to quantify the clustering of a Gaussian distributed field. It has been used to explore the substructures in the Milky Way for a long time \citep{De_Propris2010ApJ,Cooper2011MNRAS,Xue2011ApJ}. \cite{Doinidis&Beers1989ApJ} analyzed more than 4,400 BHB stars, and found an excess correlation with separations $r\leq 25$ \,pc. \cite{Starkenburg2009ApJ} applied a method of phase space correlation with 101 giants from the Spaghetti project \citep{Morrison2000AJ} to measure the amount of kinematic substructure in the Galactic halo. \cite{Mao2015AAS} measured the TPCF of G-dwarf stars within 1-3 kpc of the Sun and found some signs of small-scale substructure in the disk system of the MW. 

The stellar halo profile usually requires 3-dimensional data about the position of stars. However, the distances of stars from the Sun are usually difficult to obtain and suffer from heavy uncertainties except for some special species of stars. For example, the precision of the distance of RR Lyrae stars can be as high as $\sim 3\%$ \citep{Hernitschek2018}. However, the distance uncertainty can go up to $\sim 30\%$ for K giant stars \citep{Liu2014ApJ}. This variation introduces a significant bias to the determination of the density profile of the stellar halo. Therefore, it is of great importance to search for statistics which are immune to the measurement of distances to stars.
The ATPCF is one of the possible candidates. It has been introduced in the study of the binary stars and substructures of the MW. \cite{Lopez-Corredoira1998MNRAS} applied the ATPCF method to the Two Micron Galactic Survey catalogue to search for large-scale stellar clusters. \cite{Longhitano&Binggeli2010A&A} searched for wide binary stars from a homogeneous sample of $\sim 670 000$ main sequence stars. \cite{Lemon2004MNRAS} used this statistical technique to find substructures in the inner stellar halo ($r \lesssim 10$ kpc) and argued that there were few signs of significant structure beyond the Sagittarius (Sgr) stream. In this work, we use the ATPCF, for the first time, to measure the density profile of the MW's stellar halo. 

This paper is organized as follows: in Section \ref{sec:methods}, we describe the ATPCF and how to determine it. In Section \ref{sec:density profile}, we introduce the models used in this analysis.
The mock data used to test our method and to estimate the model prediction using ATPCF is described in Section \ref{sec:systematic}. In Section \ref{sec:fitting}, we show the test results on mock data and demonstrate the power of ATPCF in the measurement of the density profile of stars. In Section \ref{sec:practical_data}, we 
apply the ATPCF method to the observational MW data as a case study. A brief summery and conclusions are given in Section \ref{sec:conclusions}.

\section{Angular Two-Point Correlation Function}
\label{sec:methods}

\begin{figure*}
    \centering
	\includegraphics[width=0.9\textwidth]{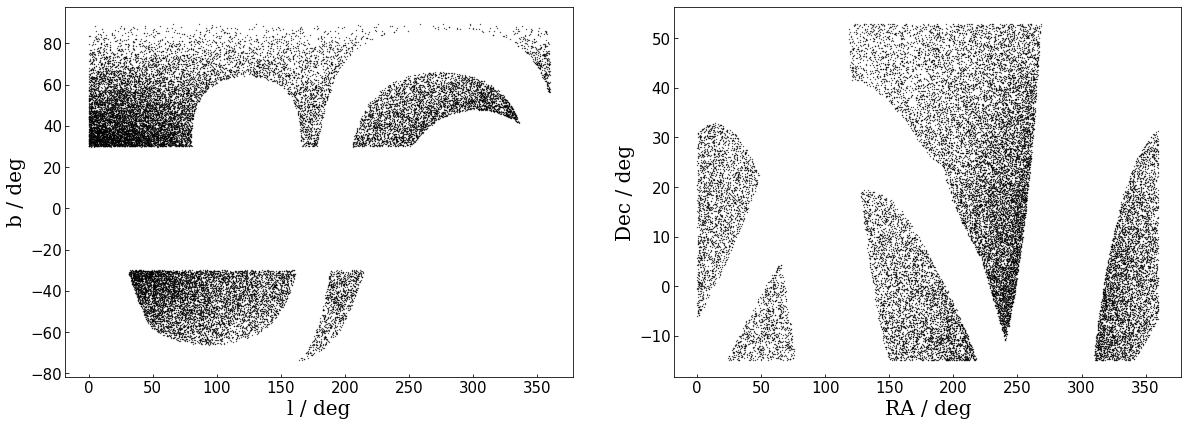} 
    \caption{The angular selection function for the fiducial survey after removing the area of the Sgr stream, and the MW's disc and bulge components.}
    \label{fig:mock_selection}
\end{figure*}

We define the angular two-point correlation function (ATPCF) by the excess in probability of finding a pair of stars separated by angular distance $\theta$ with respect to a reference background distribution.
\begin{equation}
    P = \left[1 + \omega(\theta)\right] P_{\text{BG}},
\end{equation}
where $P_{\text{BG}}$ is the probability of finding a pair of objects separated by angular distance $\theta$ in a reference background sample. To estimate the ATPCF from a stellar sample, we adopt the standard estimator 
\begin{equation}\label{ATPCF_definition}
    \omega(\theta) =\frac{DD~(\theta)}{RR~(\theta)} - 1,
\end{equation}
where $DD~(\theta)$ is the number of pairs of stars in the target sample which are separated by angular distance $\theta$, and $RR~(\theta)$ is the 
corresponding pair count of the randomly distributed points from the reference distribution.

To calculate the pair counts $DD$s and $RR$s, we use $\texttt{TreeCorr}$ \citep{Jarvis2015} which has been shown to be fast and accurate, and is efficient for our purposes. One can also choose other implementations of the correlation function, such as $\texttt{athena}$ (\citealt{Kilbinger2014ascl}), as long as the computation time is acceptable.
We bin the angular distance using equal interval binning in the log space, and take the bin interval as $\Delta \left(\ln\frac{\theta}{\rm degree}\right) = 1.2$.

A common approach is to construct a background sample with the same geometry as the data but with a larger sample size, so that the Poisson fluctuations from the reference sample can be ignored. In cosmology, a sample of uniform distribution is usually taken as the universe is homogeneous and isotropic on large scales. For the Galactic halo, one can also choose a non-uniform distribution for the background sample (e.g., \citealt{Lancaster2019}). In this work, we construct the background samples to compute $RR(\theta)$ using the same density profile as that in the mock data, but setting different parameter values to obtain a smoother distribution. For example, if the profile of the mock data sample is an SPL model with power-law index $s = 4.0$, then we also use the SPL model to generate the background sample but with $s = 1.0$. A detailed description is included in subsection ~\ref{sec:mockcf}.

\section{Density Profiles}
\label{sec:density profile}

In this section, we introduce the stellar halo density profiles used in this work. 


The coordinate system we adopt is a right-handed 3-dimensional Cartesian system with the origin at the Galactic centre. The relation between this coordinate system $(X,Y,Z)$ and the Galactic coordinate system $(l,b,D)$ is given by 
\begin{equation}\label{cart}
\begin{array}{l}
X=R_{\odot}-D \cos l \cos b \\
Y=-D \sin l \cos b \\
Z=D \sin b,
\end{array}
\end{equation}
where $l$, $b$, $D$,are the Galactic longitude, Galactic latitude, heliocentric distance respectively, and $R_{\odot}$ is the distance from Sun to the Galactic centre. We choose a fixed value for the distance from the Sun to the Galactic centre, $R_{\odot}=8.5 \text{kpc}$.  Distances  to the Galactic centre are given by
$R_\mathrm{GC}=\sqrt{X^{2}+Y^{2}+Z^{2}}$.
The perpendicular distance between the Galactic plane and the sun is less than 50 pc \citep{Karim&Mamajek2017,Iorio2018MNRAS} which is negligible for our purposes. 

The density profile of the Galactic halo could be axisymmetric, which can be described by introducing an extra parameter, the flattening factor $q$. The flattening factor is defined as the ratio of axes in the $Z$ and $X$ directions . The halo is prolate, spherical or oblate for $q>1$, $q=1$ and $q<1$, respectively. The flattening factor could vary with distance from the Galactic centre.
For non-spherical models, we redefine the radius to Galactic centre by incorporating $q$,
\begin{equation}\label{r_q}
r=\sqrt{X^{2}+Y^{2}+(Z/q)^{2}}.
\end{equation}

\subsection{SPL profile with constant flattening}
\label{subsec:SPL}

\begin{figure*}
    \centering
	\includegraphics[width=0.9\textwidth]{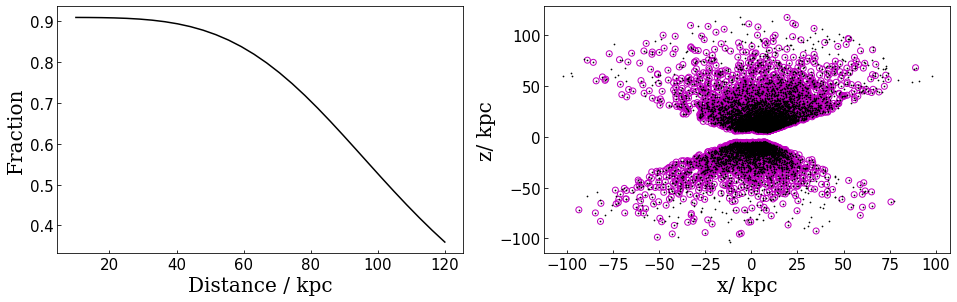} 
    \caption{Left: The radial selection function. Right: The black dots represent the mock stars draw from a SPL density profile, and the survival after applying the radial selection function are shown as pink circles.}
    \label{fig:radial_selection}
\end{figure*}
 
The density distribution of the stellar halo can be described by a single power-law profile (SPL) 
\begin{equation}
\rho(r) = \rho_{0}\left(1/r\right)^{s},
\end{equation}
where $s$ is the slope of the power-law. Here $q$ is related to the redefined radius $r$ through equation~\ref{r_q}, and is assumed to be a constant. The value of $\rho_{0}$ determines the absolute value of the number of stars in the central region. As we are only interested in the relative distribution of the stellar number density, we use just $s$ and $q$ as free parameters in our modelling. 

\subsection{SPL profile with variable flattening}
\label{subsec:SPLFF}

A varying flattening factor $q$ is generally taken as a function of Galactic radius. We use the following form for $q(R_\mathrm{GC})$ (e.g., \citealt{Xue_2015,Hernitschek2018})
\begin{equation}
q\left(R_{\mathrm{GC}}\right)=q_{\infty}-\left(q_{\infty}-q_{0}\right) \exp \left(1-\frac{\sqrt{R_{\mathrm{GC}}^{2}+r_{0}^{2}}}{r_{0}}\right),
\end{equation}
where $q_{0}$ and $q_{\infty}$ are the flattening at the Galactic centre and at a large galactocentric radius, respectively, and $r_{0}$ is the exponential scale radius influencing the degree of flattening. In this model, we have three free parameters $q_{0}$, $q_{\infty}$ and $r_{0}$. 

\subsection{BPL profile}
\label{subsec:BPL}

The third density profile utilized in this work is the broken power-law (BPL) profile which is defined by a step function as 
\begin{equation}
\rho \left(r\right)=\rho_{0}\left\{\begin{array}{l}
\left(\frac{r}{r_{0}}\right)^{-s_{1}} \text { if } D_{\min}<r<r_{0}, \\
\left(\frac{r}{r_{0}}\right)^{-s_{2}} \text { if } r_{0}<r<D_{\max},
\end{array}\right.
\end{equation}
where $s_{1}$ and $s_{2}$ control the slope of density profile in the `inner' and `outer' halo, respectively, and $r_{0}$ is the `break' radius when the slope changes from  $s_{1}$ to $s_{2}$. A constant flattening $q$ is included in $r$, and so the free parameters are $(s_{1},s_{2},r_{0},q)$.

\section{Observational effects and systematics}
\label{sec:systematic}

The ATPCF can be heavily affected by observational systematic effects such as survey geometry, or the angular and radial selection functions. It is crucial to identify and correct for these observational effects before we use the ATPCF to study the density distribution of the MW's halo.

In the following, we investigate various observational effects on the ATPCF with a set of synthetic stellar surveys of the MW. We first describe the survey geometry and angular selection effects, and then show their effects on the ATPCF using our various density distribution models. We also study the effect of unknown or unrecognised substructures by introducing globular clusters into our synthetic simulation data.

\subsection{Survey geometry and angular selection}
\label{sec:angularsf}

The geometry of a stellar survey affects the ATPCF in a rather complex way. It does not have a universal form, and varies significantly from survey to survey. Furthermore, the completeness of the sky survey is affected by many observational effects, such as weather conditions, fibre collisions, and other unexpected observational failures that occur in a real survey. The ATPCF method can be applied to both photometric and spectroscopic surveys. For the latter, one usually needs a more complicated geometrical mask on the sky.The region of sky covered may be described as “pencil beam” or “plate” like, and is usually smaller than that in a photometric survey and with a more complicated morphology.

Recognised large substructures in the stellar halo of the MW have different histories of formation and evolution from the normal halo stars, and should therefore be removed from the survey. 
The Galaxy has disk, bulge and stellar halo components. Combining these together make the analysis  difficult, especially if stellar distances are not available and we can not identify which components stars belong to. To simplify the problem, here we focus on the stellar halo, and mask any sky regions containing bulge or disk contributions. The resulting angular selection function can be formulated as
\begin{equation}
\mathcal{S}_{\text {sky}}(l,b)=\mathcal{S}_{\text {geom}}\times \mathcal{S}_{\text{sub}} \times \mathcal{S}_{\text{bulge,disc}} \times \mathcal{S}_\text{comp},
\end{equation}
where $\mathcal{S}_\text{comp}$ is the completeness of the survey and is set to be 1 in the remainder of this paper for simplicity.

We choose a simple geometry for all our synthetic surveys.  This geometry is described by $\mathcal{S}_\text{geom}$ and is given by
\begin{equation}
\mathcal{S}_{\text{geom}}(l, b)= \mathcal{S}_{\text{geom}}(\alpha, \delta)=\left\{\begin{array}{ll}
1, & \text { if } -15\degr<\delta<53\degr, \\
0, & \text { otherwise },
\end{array}\right.    
\end{equation}
where $\delta$ is the declination in the equatorial coordinate system. This geometrical cut is chosen to mimic a real survey in the northern hemisphere. The selection functions for the bulge and disc masks are combined into one component, $\mathcal{S}_\text{bulge,disc}$ which is given by
\begin{equation}
\mathcal{S}_{\text {bulge,disc }}(l,b)=\left\{\begin{array}{ll}
1, & \text { if }|b|>30\degr, \\
0, & \text { otherwise }.
\end{array}\right.    
\end{equation}
For the recognised substructures, we consider the Sagittarius (Sgr) stream which is the largest MW substructure currently detected. We adopt the definition of the Sgr stream given in \cite{Belokurov2014MNRAS}, and mask it with 
\begin{equation}
\mathcal{S}_\text{sub}(l,b)=\mathcal{S}_\text{sgr}(\tilde{\Lambda},\tilde{B})=\left\{\begin{array}{ll}
1, & \text { if }|\tilde{B}|>10\degr, \\
0, & \text { otherwise },
\end{array}\right.    
\end{equation}
where $\tilde{B}$ is one of the coordinates in the Sgr coordinate system ($\tilde{\Lambda},\tilde{B}$) introduced in \cite{Belokurov2014MNRAS}.

The resulted angular selection function together with the survey geometry is shown in Fig.~\ref{fig:mock_selection}. 

\subsection{Radial selection function}
\label{sec:radialsf}

In general, tracers of halo density distributions, such as RR Lyrae variable stars and giant stars, are complete only up to some limiting magnitude. It is usually assumed that radial incompleteness is not correlated with the angular selection function, and can be described by a radial selection function, which is the fraction of the observed stars out of the total, as a function of distance from the observer. To investigate this effect, we adopt the radial selection function for the PS1 survey \citep{Sesar2017}. As our method is more general than any specific survey, this radial selection function is taken as an example. It has the functional form
\begin{equation}
    \mathcal{S}_{\text{radial}}\left(D\right)=L-\frac{L}{1+\exp \left(-k\left(r_{\mathrm{F}}-x_{0}\right)\right)},
\end{equation}
where $r_{\mathrm{F}}=2.05 \log (D)+11$ is the extincted flux-averaged r-band stellar magnitude. 
The model parameter $L$ is the peak value of the function, $k$ is the sharpness, and $x_0$ is the magnitude at which the completeness drops to $50\%$. Here we take parameter values $L=0.91$ and $x_0=20.6$ (\citealt{Sesar2017}). The original value of parameter $k$ in \cite{Sesar2017} is $k = 4.0$. We use $k = 2.0$ to make the radial selection function sharper, and cause the fraction of stars to reduce from $\sim 40$ kpc (if $k = 4.0$ had been used, the turning point is at $\sim 70$ kpc).
The radial selection function is shown in the left panel of Fig.~\ref{fig:radial_selection}, with the right panel showing one realization of an SPL model (black solid points) selected using this selection function (pink circles).

\subsection{Residual over-densities}
\label{sec:overdensities}

In the MW, there are many substructures which do not belong to the smooth distribution of the stellar halo. These substructures are mainly globular clusters, dwarf galaxies and stellar streams. The dominant over-densities can usually be identified and eliminated. However, unrecognized substructures also have significant effects on the estimation of a smooth density profile. To show how large this effect can be, we consider one of the substructures, the globular clusters. To model their effect, we incorporate 20 mock globular clusters after having constructed the smooth halo star mock sample.  Our mock globular clusters are assumed to follow an SPL distribution radially with power index $s = 1$ and flattening $q = 1$. 
The member stars of the globular cluster are randomly sampled with total number being drawn from a Gaussian probability distribution. According to \citet{Clement2001AJ} and \citet{Hernitschek2018}, typically, the mean value and standard deviation of the total number of stars in the clusters can be set to $n_{\text{star}} = 100$ and $\sigma_{n_{\text{star}}} = 20$. The mean value of the stars in our mock globular cluster is more than the real case, which makes it easier to study the influence of the over-densities, as their signal will be stronger. 

\subsection{ATPCF on synthetic data}
\label{sec:mockcf}

\begin{figure}
	\centering
	\includegraphics[width = 0.45\textwidth]{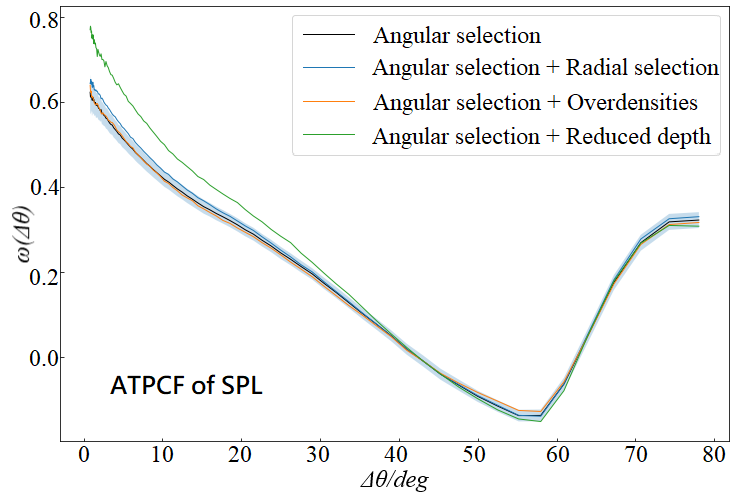}
    \caption{The ATPCFs of mock samples from the SPL density profile with constant flattening. The black curve shows the ACF of the basic model samples on which only angular selection function is applied. The blue, orange and green curves show the same ACF of mock samples with the addition of the radial selection function, globular clusters and trimmed depth respectively.}
    \label{fig:powerlaw_ATPCF}
\end{figure}

We use synthetic simulation data to study the various systematic effects, as discussed above, on the ATPCF method. For a given density distribution with specific parameters, sample data of size $N$ is constructed by the following steps.
\begin{itemize}
    \item[(1)] We sample $N'$ distances $R_i$ ($i=1,\dots,N'$), according to the radial density profile using a rejection sampling method. Here $N'$ is larger than $N$.
    \item[(2)] For each $R_i$, we draw $\phi_i$ and $Z_i$ from uniform distributions in the range [$0,2\pi$] and $[-R_i,R_i]$, respectively. The other two Cartesian coordinates $X_i$ and $Y_i$ are obtained from 
    them
        \begin{equation}
        \begin{array}{l}
        X_i=\sqrt{R_i^{2}-Z_i^{2}} \cdot \cos \phi_i, \\
        Y_i=\sqrt{R_i^{2}-Z_i^{2}} \cdot \sin \phi_i. \\
        \end{array}
        \end{equation}
     This gives us a uniform and spherically symmetric sample of mock stars.
    \item[(3)] For an oblate halo, we replace $Z_i$ with $q\cdot Z_i$ to account for the flattening. The Cartesian coordinates are then transformed to the heliocentric distance $D$, the longitude $l$ and the latitude $b$ in the Galactic coordinate system using equation~\ref{cart}.
    \item[(4)] We apply the angular selection function using the ($l, b$) coordinates of our sample.
    \item[(5)] We apply the radial selection function using the heliocentric distances of our sample.
    \item[(6)] When we study the effect of survey depth, we cut the sample as needed on heliocentric distance.
\end{itemize}

To study the various systematics, we construct different mock samples by selecting different sets of the steps described above. For example, steps 1-4 are used to get mock samples with the angular selection function only, which we refer to as the basic mock samples. While an additional step 5 is included when we study the effect of the radial selection function. A set of globular clusters as described in section \ref{sec:overdensities} are generated and placed randomly in the survey area to study the effect of residual over-densities.

\begin{figure}
	\centering
	\includegraphics[width = 0.45\textwidth]{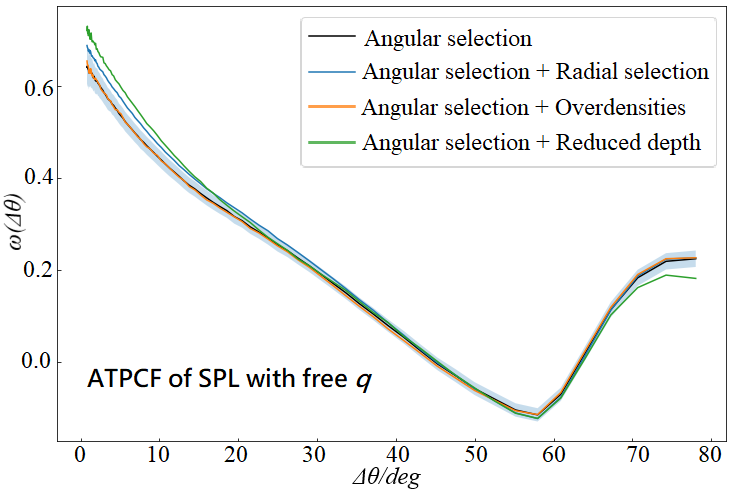}
    \caption{Same as Fig. \ref{fig:powerlaw_ATPCF} but for the SPL model with varying flattening.}
\label{fig:powerlaw_model_vflattening_ATPCF}
\end{figure}

The ATPCF is calculated using the estimator in equation \ref{ATPCF_definition}. In this study, we design smoothly distributed reference random samples for the $RR$ calculation. All the reference random samples are spherically symmetric with respect to the Galactic centre and follow the SPL distribution radially. The power index is $s=1.0$ for models with an SPL density profile and constant flattening, with $s=2.0$ for models with varying flattening. For the BPL models, we take $\{s_{1},s_{2}\} = \{1.0, 1.0\}$. The flattening factor in all of these models is $q=1.0$. 

The ATPCFs of SPL models and constant flattening factor are shown in Fig. \ref{fig:powerlaw_ATPCF}. The parameters for the density profile model are $\{s = 4.0, q = 0.9\}$. For each curve, we take the average of the correlation functions from 100 mock samples to reduce the variance. The variance for the basic mock sample are also estimated from the 100 mock samples and are shown as the shaded band. The ATPCFs of models with the radial selection function are shown as the blue solid line and are consistent with the results of the basic mock samples with at most a $1\sigma$ deviation between them.

To explore the effect of survey depth, we remove the stars from the basic mock samples whose heliocentric distances are greater than 50 kpc or lower than 10 kpc. The ATPCFs of these samples are shown as the green solid line in Fig. \ref{fig:powerlaw_ATPCF}. There are significant deviations of the ATPCFs of the depth-trimmed samples from those of the basic mock samples on angular scales of $0\degr<\Delta\theta<40\degr$. 

Also shown in Fig. \ref{fig:powerlaw_ATPCF} are the ATPCFs of the mock samples with globular clusters (the orange curve). They are consistent with those of the basic mock samples which shows that the effect of globular clusters is negligible and can be safely ignored.

\begin{figure}
	\centering
	\includegraphics[width = 0.45\textwidth]{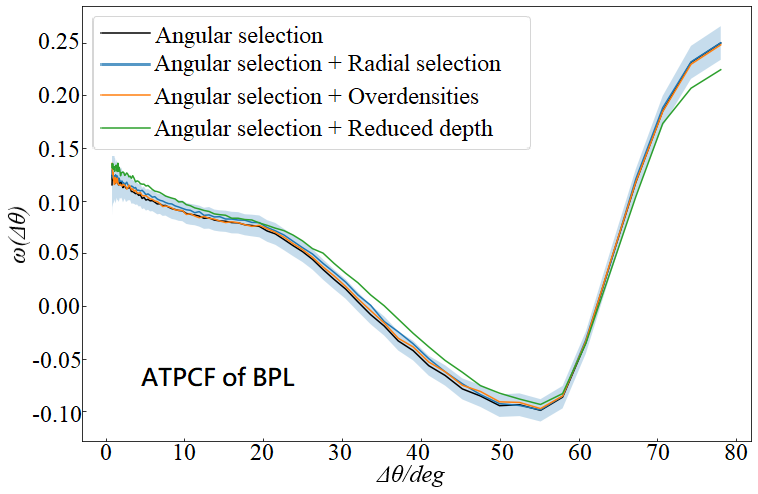}
    \caption{ Same as Fig.~\ref{fig:powerlaw_model_vflattening_ATPCF} but for the BPL model.}
    \label{fig:broken_powerlaw_model_4paras_ATPCF}
\end{figure}

For the SPL model with varying flattening and the BPL model, the ATPCFs under different observational effects are shown in Fig. \ref{fig:powerlaw_model_vflattening_ATPCF} and Fig. \ref{fig:broken_powerlaw_model_4paras_ATPCF}, respectively. For the SPL model with varying flattening, the effects of the radial selection are negligible on large angular scales and only moderate on small angular scales, and overdensities only have a negligible effect. For a reduced survey depth, the deviation of ATPCFs is not only on small scales, but also can be found on large scales. In the case of the BPL model, a reduced depth has significant impact on the ATPCF, while the radial selection has a relatively small influence, and the effect of overdensities is still negligible. The main deviation for this model shows in the middle angular scales, but there are also some deviations at large and small scales due to reduced survey depths. In our fitting process to recover the model parameters of the BPL model and the SPL model with varying flattening, we have only tested with samples constructed using both angular and radial selection functions.

\section{Recovering the halo density profile}
\label{sec:fitting}

In this section, we explore the power of the ATPCF to recover the parameters of a stellar density distribution by fitting synthetic simulation data created from the known density profiles described in the previous section. 


%
%
%
%

\subsection{Fitting Method}

\begin{table}
	\centering
	\caption{The input value of parameters and flat priors. We show the input parameters of each model and their flat priors. No matter what limitation we consider (the radial selection, survey area and depth, or over-densities), the input value of parameters and the flat priors are totally identical.}
	\label{tab:Input setting}
	\begin{tabular}{llc} 
    	\hline
                Model & Input Value&Flat Prior\\
    	\hline
    	      SPL& $s = 4.0$  & $1.0<s<6.0$\\
                &$q = 0.9$&$0.1<q<1.0$\\
            \hline
    		SPL with $q(R_\mathrm{GC})$&$s = 4.0$ & $1.0<s<5.0$ \\
                &$q_{\infty} = 0.98$&$0.1<q_{\infty}<1.0$\\
                &$q_{0} = 0.75$&$0.1<q_{0}<1.0$\\
                &$r_{0} = 20$&$0.01<r_{0}<50$\\
            \hline
    		  BPL & $s_{1} = 2.0$ & $1.0<s_{1}<6.0$ \\
                &$s_{2} = 4.0$ & $1.0<s_{2}<6.0$  \\
                &$r_{0} = 30$ & $9<r_{0}<50$ \\
                &$q = 0.8$ & $0.1<q<1.0$\\
    	\hline
	\end{tabular}
\end{table}

In order to test our approach, we need to use our ATPCF method to help fit the parameters of our density models to assess how well our method can reconstruct the input parameters of a stellar halo. We utilize Markov Chain Monte Carlo (MCMC) simulations in the fitting, and use the publicly available code $\texttt{emcee}$ \citep{Foreman-Mackey2013}, which is based on Goodman \& Weare’s affine-invariant Markov Chain Monte Carlo approach. In each step of the MCMC simulation, a sample is generated from the density profile that we chose in making the mock dataset, though we ignore some observational effects (see section~\ref{sec:systematic} for more details). The ATPCF method is applied to the simulated sample, and a $\chi^{2}$ statistic calculated,
\begin{equation}
\chi^{2} = \sum_{i} \left(\frac{\omega_{\text{mock},i}-\omega_{\text{model},i}}{\sigma_{\text{mock},i}}\right )^{2},
\end{equation}
where $\omega_\text{mock}$ and $\omega_\text{model}$  are the ATPCF values determined from the mock data sample and the sample generated in each MCMC step, respectively. The former has uncertainty estimates which are regarded as part of the input data, while the latter is computed in each simulation step and without an uncertainty estimation. The $\sigma_\text{mock}$ is the standard deviation of the ATPCF measurements of 1000 mock samples with fiducial input parameters. Here $i$ denotes each angular distance bin from which the ATPCF values are  measured, and the summation is for all angular distance bins.

The model priors are assumed to be flat, namely uniform distributions,
and are given in Table~\ref{tab:Input setting}. Since the stellar halo has a oblate shape according to  previous studies, all of the priors of $q$ are set to $q<1$. The best fitting model parameters are taken from the posterior median values obtained from the MCMC simulation. The upper and lower limits of the uncertainty are estimated using the $84.13\%$ and $15.87\%$ percentiles, respectively, corresponding to the $1 \sigma$ region for a Gaussian posterior distribution. 

\subsection{Fitting Results}
\label{sec:results}
\begin{table*}
    \centering
	\caption{The best-fitting parameters and their $1\sigma$ ($68.3 \%$) uncertainties.}
	\label{tab:uncertainties}
	\begin{tabular}{cccccc} 
		\hline
		  Model \& Parameters & Angular selection (AS) & AS \& {\color{blue} Radial selection} & AS \& {\color{blue} over-densities} & AS \& {\color{blue}Reduced area} & AS \& {\color{blue} Reduced depth}\\
		\hline
		SPL &  &  &  & & \\
		$s$ & $4.00_{-0.04}^{+0.04}$ & $4.00_{-0.04}^{+0.04}$ & $3.96_{-0.04}^{+0.04}$ & $3.99_{-0.06}^{+0.06}$ & $4.00_{-0.05}^{+0.05}$ \\
		$q$ & $0.90_{-0.02}^{+0.02}$ & $0.90_{-0.02}^{+0.02}$ & $0.91_{-0.02}^{+0.02}$ & $0.90_{-0.04}^{+0.04}$ & $0.90_{-0.02}^{+0.02}$ \\
		\hline
		SPL with $q(R_\mathrm{GC})$  &  &  &  &  & \\
		$s$ & $4.01_{-0.04}^{+0.04}$ & $4.01_{-0.04}^{+0.05}$ &  & \\
		$q_{\infty}$  & $0.98_{-0.02}^{+0.02}$ & $0.99_{-0.02}^{+0.01}$ &  & \\
		$q_{0}$  & $0.63_{-0.36}^{+0.25}$ & $0.63_{-0.34}^{+0.26}$ &  & \\
		$r_{0}$  & $22.09_{-14.81}^{+17.24}$ & $23.23_{-15.50}^{+16.20}$ &  & \\
		\hline
		BPL &  &  &  & \\
		$s_{1}$ & $2.12_{-0.53}^{+0.48}$ & $2.11_{-0.58}^{+0.50}$ &  \\
        $s_{2}$ & $3.93_{-1.03}^{+1.22}$ & $4.02_{-1.09}^{+1.24}$ &  \\
		$r_{0}$ & $31.77_{-10.16}^{+10.07}$ & $31.33_{-10.22}^{+11.02}$ &  \\
		$q$ & $0.80_{-0.03}^{+0.02}$ & $0.81_{-0.03}^{+0.03}$ &  \\
		\hline
	
	\end{tabular}
\end{table*}

We now discuss the results from fitting the synthetic data generated from the models described in Section \ref{sec:density profile}. The 1- and 2-dimensional marginalized distributions of the parameters for different models are plotted using the publicly available $\texttt{corner}$ code \citep{corner}, which is a Python module to visualize multidimensional samples by means of a scatter plot matrix. For the SPL model, as a typical and simple case, we investigate not only the angular selection and radial selection functions as constraints, but also the impact of the over-densities, and the area and depth of the sample. The results are given in Table~\ref{tab:uncertainties}. For the other two models, we only examine the angular selection and  radial selection effects. 

\begin{figure}
	\centering
	\includegraphics[width = 0.35\textwidth]{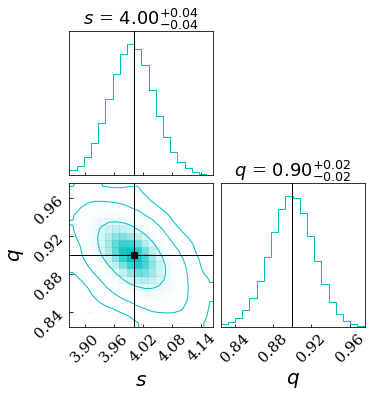}
    \caption{ One-dimensional histograms (top left and bottom right) and two dimensional projected contours (bottom left) of the posterior probability distributions of the parameters ($s$, $q$) of the SPL density profile with constant flattening fitting to the mock data after considering the angular selection. The black cross lines indicate the best fitting, and the $1\sigma$ intervals are recorded on the top of the histograms. }
    \label{fig:powerlaw_no_radial_selection_sim}
\end{figure}
\begin{figure}
	\centering
	\includegraphics[width=0.35\textwidth]{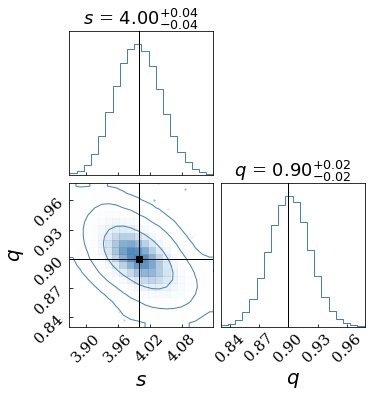}
    \caption{Similar to ~\ref{fig:powerlaw_no_radial_selection_sim}, the only difference is that both the angular and radial selections are considered here.}
    \label{fig:powerlaw_radial_selection_sim}
\end{figure}
\begin{figure}
    \centering
	\includegraphics[width=0.45\textwidth]{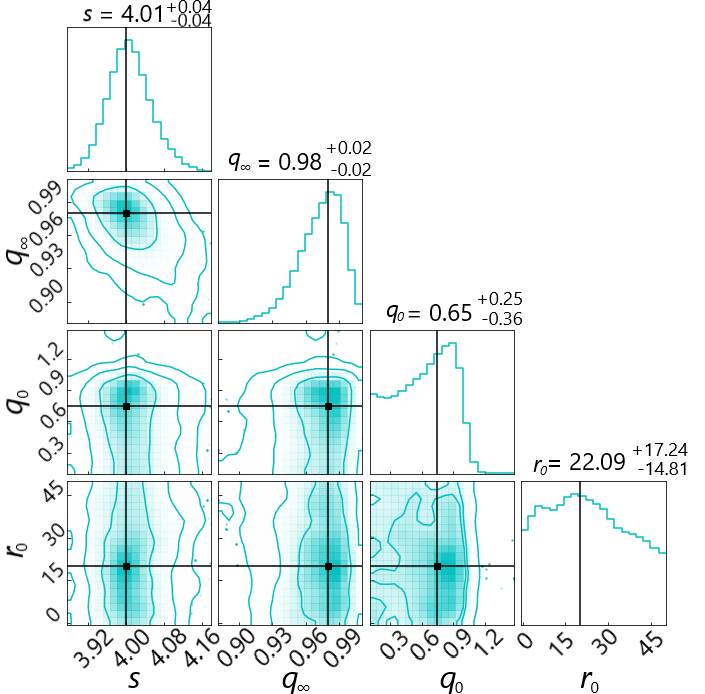}
    \caption{ One-dimensional histograms and two dimensional projected contours of the posterior probability distributions of the parameters ($s$, $q_{0}$, $q_{\infty}$ and $r_{0}$ ) of the SPL density profile with varying flattening to fit the mock data with only the angular selection. The black cross lines indicate the best fitting, and the $1\sigma$ intervals are recorded on the top of the histograms.}
    \label{fig:powerlaw_vflattening_no_radial_selection_sim}
\end{figure}
\begin{figure}
    \centering
	\includegraphics[width=0.45\textwidth]{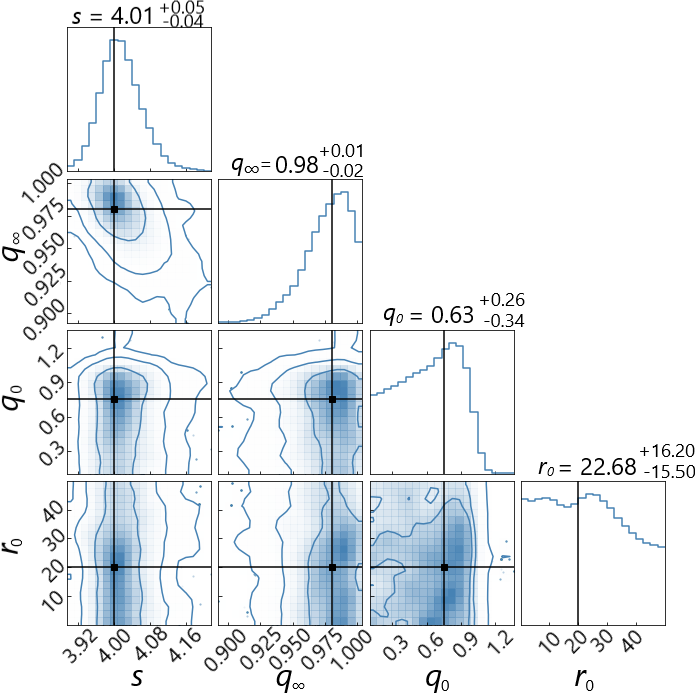}
    \caption{Similar to Fig.~\ref{fig:powerlaw_vflattening_no_radial_selection_sim}, the only difference is that both the angular and radial selections are considered here.}
    \label{fig:powerlaw_vflattening_radial_selection_sim}
\end{figure}
\begin{figure}
    \centering
	\includegraphics[width=0.45\textwidth]{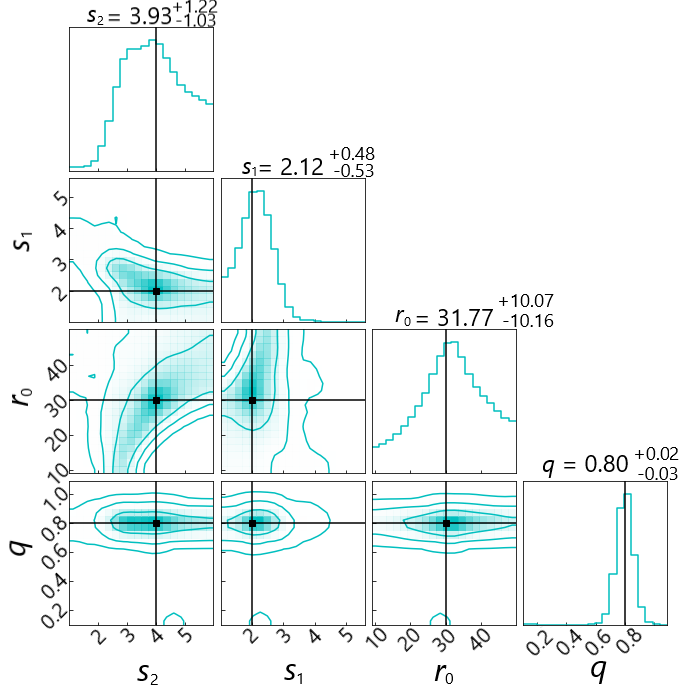} 
    \caption{ One-dimensional histogram and two dimensional projected contours of the posterior probability distributions of the parameters ($s_{1}$, $s_{2}$, $q$ and $r_{0}$ ) of the BPL density profile fitted to the mock data with only the angular selection. The black cross lines indicate the best fitting, and the $1\sigma$ intervals are recorded on the top of the histograms.}
    \label{fig:broken_powerlaw_4paras_no_radial_selection_sim}
\end{figure}
\begin{figure}
    \centering
	\includegraphics[width=0.45\textwidth]{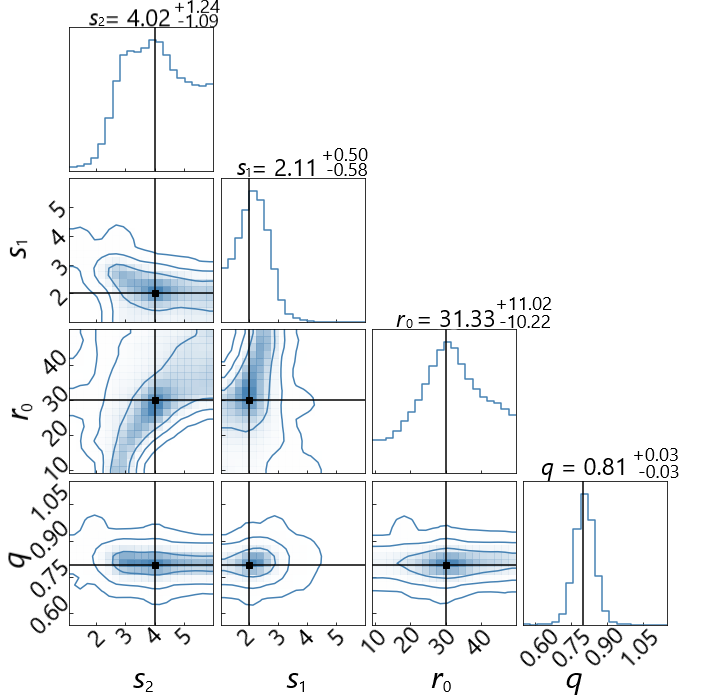} 
    \caption{Similar to Fig.~\ref{fig:broken_powerlaw_4paras_no_radial_selection_sim}, the only difference is that both the angular and radial selections are considered here.}
    \label{fig:broken_powerlaw_4paras_radial_selection_sim}
\end{figure}

\begin{enumerate}
\item As shown in Fig.~\ref{fig:powerlaw_ATPCF}, the ATPCF of the SPL model is not much affected when we vary angular selection, radial selection and the over-densities. However, the difference in the ATPCFs is more than $1\sigma$ when the radii of the stars are limited to $0-50$ kpc compared to the samples with an upper distance limit of $120$ kpc.  
We can see in Table~\ref{tab:uncertainties} that the parameters of the SPL model can be recovered accurately and they are not affected by the various observational effects we considered.
The radial selection function and the globular substructures have negligible impact on the parameters. 
However, with a smaller survey depth or area, the uncertainties become large, which is consistent with the result shown in Fig.~\ref{fig:powerlaw_ATPCF}. The marginalized distributions of the recovered $s$ and $q$ when considering angular selection with and without radial selection are given in Fig.~\ref{fig:powerlaw_no_radial_selection_sim} and Fig.~\ref{fig:powerlaw_radial_selection_sim}. 

\item When only angular selection is considered and using a variable flattening factor, we have 4 parameters in the SPL with $q(R_\mathrm{GC})$, and find that $s = 4.01_{-0.04}^{+0.04}$, $q_{\infty} = 0.98_{-0.02}^{+0.02}$, $q_{0} = 0.65_{-0.36}^{+0.25}$ and $r_{0} = 22.09_{-14.81}^{+17.24}$. After including radial selection as well, we have $s = 4.01_{-0.04}^{+0.05}$, $q_{\infty} = 0.98_{-0.02}^{+0.01}$, $q_{0} = 0.63_{-0.34}^{+0.26}$ and $r_{0} = 22.68_{-15.50}^{+16.20}$. The marginalized distributions of the parameters are shown in Fig.~\ref{fig:powerlaw_vflattening_no_radial_selection_sim} and \ref{fig:powerlaw_vflattening_radial_selection_sim}. The results show that $s$ and $q_{\infty}$ are well constrained, and the other two parameters less so.

\item For the BPL model, our results are not as good as studies that used full 3D positional data (e.g., \citealt{Hernitschek2018}). We have $s_{1}=2.12_{-0.53}^{+0.48}$, $s_{2}=3.93_{-1.03}^{+1.22}$, $r_{0}=31.77_{-10.16}^{+10.07}$, $q=0.80_{-0.03}^{+0.02}$ without radial selection, and $s_{1}=2.11_{-0.58}^{+0.50}$, $s_{2}=4.02_{-1.09}^{+1.24}$, $r_{0}=31.33_{-10.22}^{+11.02}$, $q=0.81_{-0.03}^{+0.03}$ with radial selection included. The marginalized distributions of the recovered parameters are indicated in Fig.~\ref{fig:broken_powerlaw_4paras_no_radial_selection_sim} and \ref{fig:broken_powerlaw_4paras_radial_selection_sim}. These results are still useful if one investigates the constraints on the BPL model. For example, the best fitting BPL parameter values for $r_{0}$ in \cite{Deason2011}, \cite{Xue_2015} and \cite{Hernitschek2018} are $27$, $18$ and $38$ kpc, respectively. Their differences are at the same order of magnitude as the upper and lower limits of our recovered $r_{0}$, and the best fitting values often have some discrepancies which can not be ignored when different data samples are utilized. Furthermore, we obtain some evidence that the Galactic halo is oblate by determining its ATPCF from real data if we assume that the halo profile follows the BPL model (see the section~\ref{sec:practical_data}).
\end{enumerate}

Overall, these results indicate that our method can reconstruct the smooth component of a stellar halo using only 2D angular position, provided that the maximum and minimum distances of the data sample are included in the modelling, and provided that a specific number density model is assumed. The minimum and maximum radial distances are required, particularly in the parametric models that contain distance information, e.g., $r_{0}$ in the BPL model indicating the location of the 'break': the lower and upper limits of the distances in the stellar sample have a normalization effect. Moreover, parameter uncertainties are model dependent with our approach, with the SPL model having the lowest uncertainties among the models we choose. 

\section{Application to real data}
\label{sec:practical_data}

Finally, we apply the ATPCF to real MW data. The tracers in the data set are type ab RR Lyrae stars (RRLab) from the Catalina Surveys Data Release 1 \citep{Drake2014ApJS}. This data set has already been employed by \cite{Ablimit2018} when they used a 3-parameter BPL model to fit the halo's stellar distribution. The parameters in their model are $(s_1,s_2,r_0)$, which means that the flattening factor in our 4-parameter BPL model should be fixed to 1.0. To compare our results with some previous studies (e.g., \citealt{Watkins2009}, \citealt{Faccioli2014ApJ} and \citealt{Ablimit2018}), we first use this 3-parameter BPL profile.

\begin{figure}
	\centering
	\includegraphics[width=0.39\textwidth]{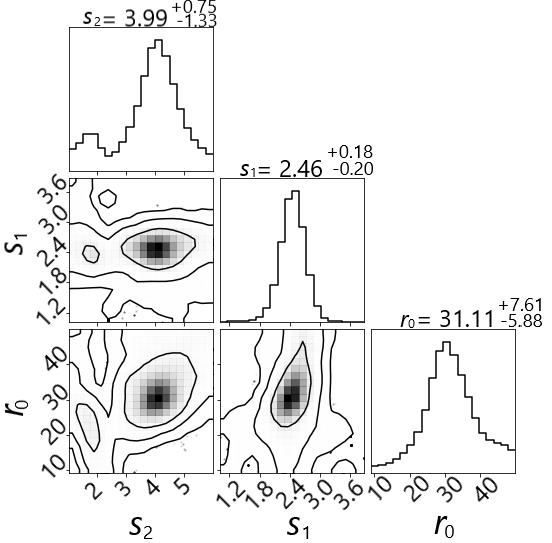}
    \caption{ One-dimensional histogram and two dimensional projected contours of the posterior probability distributions of the parameters ($s_{1}$,$s_{2}$, $r_{0}$) of the BPL density profile without flattening factor, fitted to RRLab data from Catalina survey after removing the Sgr stream. The $1\sigma$ intervals are recorded on the top of the histograms.}
    \label{fig:real_data_broken_powerlaw_without_sgr}
\end{figure}

There are 12,227 type ab RR Lyrae in the data set. We limit the radius in the range of 9-50 kpc, which is the same as that in \cite{Ablimit2018}. The angular selection is the same as in Section~\ref{sec:systematic}, which is actually designed for this data set to reduce the influence caused by the incompleteness of the edges.
Here, of course, we replace the mock data with Catalina RRLab data. The impact of the Sgr stream has been discussed in several studies (e.g., \citealt{Drake2013ApJ,Faccioli2014ApJ,Ablimit2018}), and it is also considered in our work. Although \cite{Drake2013ApJ} claimed that the Sgr stream has an important impact on their fitting results, \cite{Faccioli2014ApJ} and \cite{Ablimit2018} both supported ignoring the stream. Our sample has 7013 stars if we preserve the Sgr stream, and 4791 stars if we remove it. We use the same method as in section~\ref{sec:systematic}, and the background sample is identical to the BPL model in section~\ref{sec:density profile}. Our best fitting parameters are $s_{1}=2.42_{-0.23}^{+0.21}, s_{2}=4.35_{-1.43}^{+0.87}, r_{0}=31.23_{-8.38}^{+7.84}$ if we include the Sgr stream. After removing it, we have $s_{1}=2.46_{-0.20}^{+0.18}, s_{2}=3.99_{-1.33}^{+0.75}, r_{0}=31.11_{-5.88}^{+7.61}$. We conclude therefore that the difference between keeping and eliminating the Sgr stream is very small for this model. The resulting contour plot is shown in Fig.~\ref{fig:real_data_broken_powerlaw_without_sgr}. 

\cite{Watkins2009} used RRLyrae from the Sloan Digital Sky Survey Stripe 82, and obtained $\left(s_1, s_2, r_{0}\right)=\left(2.4,4.5,23 \mathrm{kpc}\right)$. \cite{Faccioli2014ApJ} fitted the density profile using 318 RRLs observed by the Xuyi Schmidt telescope photometric survey, and determined $s_1 = 2.3 \pm 0.5$, $s_2 = 4.8 \pm 0.5$, and $r_0 =21.5 \pm 2.2 \mathrm{kpc}$. Using 860 RRLab stars as halo tracers, as in \cite{Drake2013ApJ}, combined with LAMOST DR4 and PDSS DR8 data, \cite{Ablimit2018} found the BPL parameters  were $s_1=2.8 \pm 0.4$, $s_2=4.8 \pm 0.4$ and $r_{0}=21 \pm 2 \mathrm{kpc}$. \cite{Medina2018ApJ} utilized 173 RR Lyrae stars over $\sim 120 \mathrm{deg}^{2}$, with heliocentric distances of the full sample range from 9 to $>200$ kpc, and found a power-law index value of $s=-4.17_{-0.20}^{+0.18}$ for their SPL model. \cite{Fukushima2019PASJ} used BHB stars with the Galactocentric radius ranging from 36 to 360 kpc, and found that they can be approximated as an SPL profile with index $s=-3.74_{-0.21}^{+0.22}$. \cite{Stringer2021ApJ} used RR Lyrae stars from the full six-year data set of the Dark Energy Survey, which covers $\sim 5000$ $\deg^2$ of the southern sky. They detected 6971 RR Lyrae candidates out to $\sim 335$ kpc, with $>70 \%$ completeness at $\sim 150$ kpc, and found that $s_{1}=-2.54_{-0.09}^{+0.09}$, $s_{2}=-5.42_{-0.14}^{+0.13}$ and $r_{0}=32.1_{-0.9}^{+1.1}$ as a result of fitting the BPL model with flattening $q$ fixed to 0.7. Our results are consistent with these studies.

\begin{figure}
	\centering
	\includegraphics[width = 0.45\textwidth]{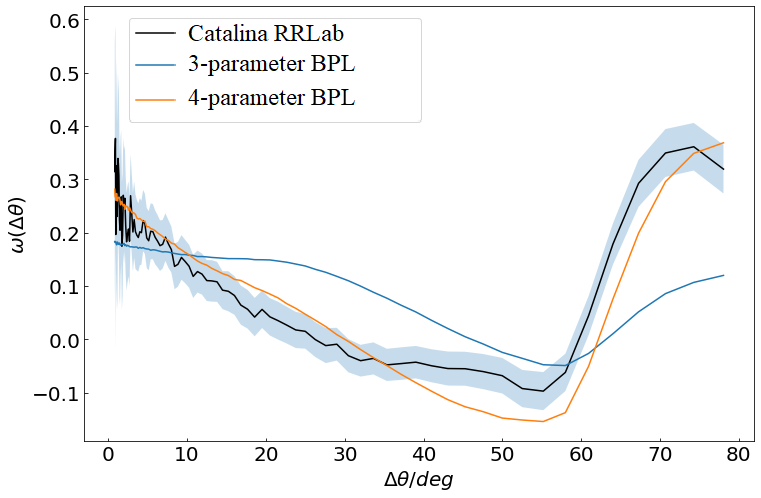}
    \caption{ ATPCF of real data and two models. The black filled circles are ATPCF of RRLab from Catalina survey. Blue and orange lines are ATPCFs measured from 3-parameter and 4-parameter BPL model, respectively.}
    \label{fig:real_data_ATPCF}
\end{figure}

However, in our work, the ATPCF computed from the best fitting 3-parameter BPL model is not consistent with that calculated from the data. For this reason, we use the 4-parameter BPL model (which is introduced in section~\ref{sec:density profile}) to fit the RRLab data. The additional parameter in this model is the flattening $q$. A comparison of ATPCFs determined from RRLab data and the best fitting 3- and 4-parameter BPL models is shown in Fig.~\ref{fig:real_data_ATPCF}. It tells us that the ATPCF from 4-parameter BPL model is closer to that from the real data. The flattening factor is a reasonable parameter to be included in the halo's profile, with an oblate stellar halo being supported by \cite{Deason2011,Sesar2011,Pila-Diez2015,Xu2018,Hernitschek2018,Stringer2021ApJ}. Therefore, although a model with more parameters might possibly fit the data better, this result implies that the MW's stellar halo is oblate.

\section{Conclusions}
\label{sec:conclusions}

\begin{figure}
	\centering
	\includegraphics[width=0.49\textwidth]{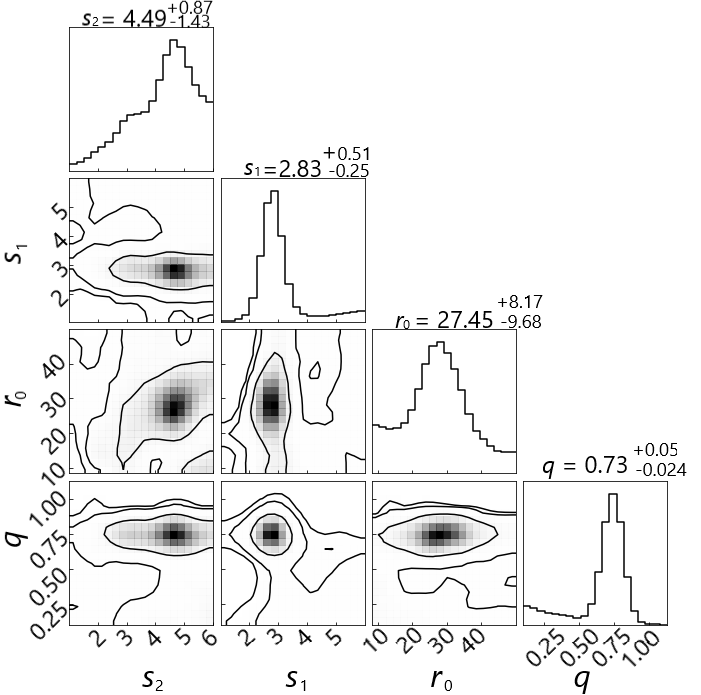}
    \caption{ One-dimensional histogram and two dimensional projected contours of the posterior probability distributions of the parameters ($s_{1}$,$s_{2}$, $r_{0}$,$q$) of the BPL density profile, fitted to RRLab data from Catalina survey after removing the Sgr stream. The $1\sigma$ intervals are recorded on the top of the histograms.}
    \label{fig:broken_powerlaw_realdata_without_sgr_4para}
\end{figure}

We have developed an approach to determining the number density distribution of stars within the Galactic halo using 2-dimensional stellar spatial data. Our approach is based on computing the ATPCF of the stars' positional distribution projected onto the celestial sphere. The ATPCF for the real data is compared with the ATPCFs of simulated halos produced from parameterized density profiles to understand which profile and parameter values best represent the MW halo.  Our rationale for using 2-dimensional data is to eliminate the potentially large uncertainties arising from distance measurements to MW stars.

Two generally well-known, parameterized, stellar density profiles of the MW halo are utilized, the SPL and BPL profiles. Both profiles use a flattening factor $q$ to help describe the shape of the halo. For the SPL model, we introduce a radially varying flattening factor $q(R_\mathrm{GC})$ and investigate its effect. 

We have illustrated the effects caused by the uncertainties in distances by generating some mock data sets from the density profiles, and measuring the errors by averaging 1000 mock samples made from the models with fiducial parameters. 
To investigate the impact of the observational selection effects, we have applied a typical angular selection to a typical sky area from which Galactic disk and bulge stars have been excluded. Substructures such as the Sgr stream are also excluded.  The 'observational area' has been limited to $-15\degr<\delta<53\degr$ to reduce the effects caused by the incompleteness in the RRLab sample.
This geometrical cut thus attempts to mimic a real survey in the northern hemisphere. 
Even though we wish to eliminate distance uncertainties from our modelling, we have sought to ensure that 
the observed data and the simulated data have similar distance limits applied to the data sets.  Based on the real data, we have applied minimum and maximum distances of $10$ and $120$ kpc to the simulated data.
The impact of the radial selection function has been taken into account in our modelling. Further more, we have discussed more complicated situations, and have taken the influence of over-densities, survey area and survey depth into account as well.

To prove the feasibility of the method, we have performed Markov Chain Monte Carlo (MCMC) parameter determinations on mock data sets generated from different density profiles.  For SPL models, we have established a good parameter determination with low uncertainties using only angular selection.  The uncertainties were similar when we added in other effects, but weakened when the survey depth and area are smaller.  When using a varying flattening, out of the four parameters involved, we found that the power law slope and the flattening at large radius were well constrained but the other two less so.
For the BPL model, we found that the inner slope and the flattening were well determined with low uncertainties but the other parameters had larger uncertainties.  The results from other studies do have significant variations, perhaps as a result of different tracers or data processing. Therefore our parameter determinations with larger uncertainties (like $r_{0}$) may still be useful. 
For the SPL model with varying flattening and the BPL model, the radial selection function we chose did not have an obvious impact on our results.

We have applied our approach to real MW data using the RRLab tracers from Catalina Survey Data Release 1. We first used a 3-parameter BPL model to fit the data. We found that the difference between keeping and removing the Sgr stream was very small. The parameter values we determined are $s_{1}=2.42_{-0.23}^{+0.21}, s_{2}=4.35_{-1.43}^{+0.87}$ and $r_{0}=31.23_{-8.38}^{+7.84}$. These results agree with \cite{Watkins2009}, \cite{Faccioli2014ApJ} and \cite{Ablimit2018}, which also utilized RR Lyrae stars as halo tracers. 
Nevertheless, the ATPCF computed from the best-fitting 3-parameter BPL profile still had relatively large deviations from that determined directly from the data. Refitting the data with a 4-parameter BPL profile, which includes the flattening factor $q$, the parameter values  are $s_{1}=2.83_{-0.25}^{+0.51}$, $s_{2}=4.49_{-1.43}^{+0.87}$, $r_{0}=27.45_{-9.68}^{+8.17}$, $q=0.73_{-0.24}^{+0.05}$. 
More importantly, the resulting ATPCF was a better match to the ATPCF computed from the real MW data implying that the MW has an oblate stellar halo.

We believe our method has the potential to make stars that lack precise distant measurements usable as tracers. Moreover, our method can not only be applied to study the stellar halo, but can also be generalized to investigate other components of the Milky Way.


\section*{Acknowledgements}
We thank Chao Liu and Hao Tian for useful discussions. 
We acknowledge support from the National SKA Program of China (No.2022SKA0110100), and the CAS Interdisciplinary Innovation Team (JCTD-2019-05). We acknowledge science research grants from the China Manned Space Project (No.CMS-CSST-2021-B01).

\section*{Data Availability}
The data underlying this article will be shared after a reasonable request to the corresponding author.


\bibliographystyle{mnras}
\bibliography{bibliography}






\bsp	
\label{lastpage}

\end{document}